\def \SAIT #1 #2 {{\em Mem.\ Soc.\ Astron.\ It.\/} {\bf #1}, #2}
\def \MESS #1 #2 {{\em The Messenger\/} {\bf #1}, #2}
\def \ASTRNACH #1 #2 {{\em Astron. Nach.\/} {\bf #1}, #2}
\def \AAP #1 #2 {{\em Astron. Astrophys.\/} {\bf #1}, #2}
\def \AAL #1 #2 {{\em Astron. Astrophys. Lett.\/} {\bf #1}, L#2}
\def \AAR #1 #2 {{\em Astron. Astrophys. Rev.\/} {\bf #1}, #2}
\def \AAS #1 #2 {{\em Astron. Astrophys. Suppl. Ser.\/} {\bf #1}, #2}
\def \AJ #1 #2 {{\em Astron. J.\/} {\bf #1}, #2}
\def \ANNREV #1 #2 {{\em Ann. Rev. Astron. Astrophys.\/} {\bf #1}, #2}
\def \APJ #1 #2 {{\em Astrophys. J.\/} {\bf #1}, #2}
\def \APJL #1 #2 {{\em Astrophys. J. Lett.\/} {\bf #1}, L#2}
\def \APJS #1 #2 {{\em Astrophys. J. Suppl.\/} {\bf #1}, #2}
\def \APSS #1 #2 {{\em Astrophys. Space Sci.\/} {\bf #1}, #2}
\def \ASR #1 #2 {{\em Adv. Space Res.\/} {\bf #1}, #2}
\def \BAIC #1 #2 {{\em Bull. Astron. Inst. Czechosl.\/} {\bf #1}, #2}
\def \JSQRT #1 #2 {{\em J. Quant. Spectrosc. Radiat. Transfer\/} {\bf #1}, #2}
\def \MN #1 #2 {{\em Mon. Not. R. Astr. Soc.\/} {\bf #1}, #2}
\def \MEM #1 #2 {{\em Mem. R. Astr. Soc.\/} {\bf #1}, #2}
\def \PLR #1 #2 {{\em Phys. Lett. Rev.\/} {\bf #1}, #2}
\def \PASJ #1 #2 {{\em Publ. Astron. Soc. Japan\/} {\bf #1}, #2}
\def \PASA #1 #2 {{\em Publ. Astron. Soc. Australia\/} {\bf #1}, #2}
\def \PASP #1 #2 {{\em Publ. Astr. Soc. Pacific\/} {\bf #1}, #2}
\def \NAT #1 #2 {{\em Nature\/} {\bf #1}, #2}
\def\proptosima{$\; \buildrel \propto \over \sim \;$}
\def\proptosim{\lower.5ex\hbox{\proptosima}}            
\def\ltsima{$\; \buildrel < \over \sim \;$}
\def\simlt{\lower.5ex\hbox{\ltsima}}            
\def\gtsima{$\; \buildrel > \over \sim \;$}
\def\simgt{\lower.5ex\hbox{\gtsima}}            

\def\ga{\mathrel{\hbox{\rlap{\hbox{\lower4pt\hbox{$\sim$}}}\hbox{$>$}}}}
\def\la{\mathrel{\hbox{\rlap{\hbox{\lower4pt\hbox{$\sim$}}}\hbox{$<$}}}}

\documentstyle{memsait}
\input epsf.sty
\begin{opening}
\title{AGN IN THE MULTIMISSION ARCHIVE AT STSCI}
\author{Paolo Padovani$^{1,2,3}$, Tim Kimball${^1}$} 
\institute{$^1$Space Telescope Science Institute, 3700 San Martin Drive, 
Baltimore MD. 21218, USA\\
$^2$Affiliated to the Astrophysics Division, Space Science Department, European
Space Agency\\
$^3$On leave from Dipartimento di Fisica, II Universit\`a di Roma 
``Tor Vergata'', Italy}
\date{} 
\end{opening}

\begin{document}

\oddpagefooter{}{}{} 
\evenpagefooter{}{}{} 
\ 
\bigskip

\begin{abstract}
We describe a new WWW interface that allows the cross-correlation of an Active
Galactic Nuclei catalog with various archives (HST, IUE, EUVE) available at
the Multimission Archive at the Space Telescope Science Institute. Details
of the catalog and the interface are provided. 
\end{abstract}

\section{The Multimission Archive at the Space Telescope Science Institute }
The Hubble Data Archive (HDA), located at the Space Telescope Science
Institute (STScI), contains, as of July 1998, over 5 Terabytes of science and
engineering data, for a total of approximately 150,000 science
exposures. Based on the success of the HDA, and taking advantage of its
existing archive infrastructure, the STScI archive has recently expanded by
providing access to non-HST datasets. The Multimission Archive at the Space
Telescope Science Institute (MAST) was thus created. This includes, at the
time of writing, the International Ultraviolet Explorer (IUE) Final Archive,
the Copernicus (OAO-3) Archive, the Extreme Ultraviolet Explorer (EUVE)
Archive, the Ultraviolet Imaging Telescope (UIT) Archive, the Faint Images of
the Radio Sky at Twenty-centimeters (FIRST) Archive, and the Digitized Sky
Survey (DSS). STScI plans to incorporate additional ultraviolet and optical
data sets into MAST in the future, including data from the Far Ultraviolet
Spectroscopic Explorer (FUSE) currently scheduled for launch in early 1999.
The MAST WWW interface is at {\tt http://archive.stsci.edu/mast.html}.

\section{AGN in MAST}
As a first step towards taking advantage of various archives at one site and
enhancing the potential of the single archives, we have started a project
which allows the cross-correlation of astronomical catalogs with the archives
available at the MAST. This interface, available on the Web at {\tt
http://archive.stsci.edu/search}, enables, as of April 1998, the
cross-correlation of an AGN catalog with the HST, IUE, and EUVE archives.

The AGN catalog, heavily based on the V\`eron-Cetty \& V\`eron (1996) [VV96]
catalog, is discussed and used for astrophysical applications by Padovani et
al. (1997) and Padovani (1998). The VV96 catalog includes 11,442 quasars and
active galaxies, and gives optical magnitudes, redshift, and some radio
information. To this Padovani et al. have added: 1. the BL Lac catalog of
Padovani \& Giommi (1995), updated with BL Lacs discovered in 1996 (for a
total of 265 sources); 2. the radio galaxies in the 1 Jy, S4, and S5 radio
catalogs, mostly not included in VV96. The resulting database, which totals
12,021 AGN, was also cross-correlated with various radio catalogs providing 6
cm data, namely the PKS database, the PMN survey , the GB6 catalog, the 1 Jy,
S4, and S5 radio catalogues. Individual radio fluxes for radio-quiet AGN not
included in radio catalogs (radio fluxes $< 1 - 30$ mJy), taken from the
literature, were also added. The V magnitudes in VV96 are actually mostly B or
photographic magnitudes when no ($B-V$) value is available. Therefore, for
objects without ($B-V$) colors, V magnitudes have been derived from the given
values by subtracting ($B-V$) values typical of the class to which an object
belongs to, unless the reference was to a paper which gave V magnitudes
directly. 

Great care has been taken in the classification of the sources. This is mostly
based on the one given by VV96 with some important differences and
additions. Namely, a distinction is made between radio-loud and radio-quiet
quasars (based on the value of the two point radio-optical spectral index
$\alpha_{\rm ro}$, 0.19 being the dividing line [this corresponds to a
(standard) dividing value of the ratio of radio flux to optical flux of
10]). Also, the ``radio galaxy'' class has been introduced and radio-loud
Seyfert 1 and 2 have been included with the radio-loud quasars and radio
galaxies respectively.

Using this interface, one can select AGN by class, redshift, magnitude, and 6
cm radio flux and cross-correlate them with the HST, IUE, and EUVE archives.
Ranges of parameters can be provided (for example, one can select all AGN with
$3 < z < 4$ and $V > 20$). For HST, one can select individual instruments,
each with a different correlation radius. Multiple missions can also be
selected, with the option to show only those AGN that cross-correlate with
every selected mission (so one can look for AGN that have been observed with
both HST and IUE, or for AGN observed with either HST or IUE). After the
correlation is performed the user is presented with a list of matches and one
can preview the images/spectra (at present only in the case of HST data), and
retrieve the data. We are currently working on the addition of the FIRST
catalog to the list of archives that can be cross-correlated with the AGN
catalog.

A user supplied catalog can also be cross-correlated with any of the above
archives. We are planning to expand this interface by including cluster,
galaxy, and stellar catalogs. MAST is supported by NASA under a cooperative
agreement between STScI and Goddard Space Flight Center.




\end{document}